\documentstyle[twocolumn,aps,epsfig]{revtex}
\begin{document}
\draft
\twocolumn[\hsize\textwidth\columnwidth\hsize\csname @twocolumnfalse\endcsname
\title{Role of the superexchange interaction in magnetic transition
and polaron crossover}
\author{ Jayita Chatterjee$^{1}$ and A. N. Das$^{2}$}
\address{ $^1${\it Department of Physics,
           Pohang University of Science and Technology,
           Pohang 790-784, Korea. \\
          $^2$Theoretical Condensed Matter Physics Division,
              Saha Institute of Nuclear Physics,\\
              1/AF Bidhannagar, Kolkata 700064, India.}           
}
\date{\today}
\maketitle

\begin{abstract}
 The Hubbard-Holstein model is studied including double-exchange
interaction and superexchange interaction using 
a variational phonon basis obtained 
through the modified Lang-Firsov (MLF) transformation followed
by the squeezing transformation. 
The kinetic energy, polaron crossover and magnetic transition
are investigated as a function of electron-phonon 
($e$-ph) coupling and electron concentration for different
values of antiferromagnetic superexchange interaction ($J$)
between the core spins.
The polaron crossover, magnetic transition and the
suppression of ferromagnetic transition with $J $ are discussed
for the model.
\end{abstract}

\vskip2pc]
\narrowtext
Recently the interest in the double exchange model \cite{zen}
has grown considerably with the discovery of very large negative 
colossal magnetoresistance (CMR) \cite{cm} 
and anomalous magnetotransport properties in doped manganites\cite{von}, 
namely in R$_{1-x}$A$_{x}$MnO$_{3}$ (where R=La, Pr and A=Ca, Sr, Ba).
Ferromagnetism in these compounds (for
$x$$\sim$ 0.2-0.4) is believed to be due to the `double-exchange' (DE)
mechanism which
operates when local Mn-ion spins
are strongly coupled by Hund's rule with the spins of
the itinerant electrons occupying a narrow band.
However, the experimental results \cite{von} in manganites, namely 
the sharp change in resistivity near $T_c$ and the physics of CMR, cannot be  
explained by the DE alone\cite{M}. Millis suggested
the lattice polaron effects due to strong electron-phonon
($e$-ph) interaction as a necessary extension\cite{Millis}.
R\"{o}der $et~ al.$ \cite{rod} also examined the combined effect of $e$-ph
interaction and DE on $T_c$ using the variational wave function techniques. 
In fact, the contribution of the lattice polaron to carrier mobility
was pointed out earlier by Goodenough \cite{good}.
Several theoretical models have been proposed based on lattice-carrier
coupling \cite{rod,lee,yu2,Dg,yar}. Many experiments
\cite{exp} indicate evidence of strong lattice-electron coupling
in Manganese Oxides \cite{exp} which shapes
the properties of manganites crucially. Moreover, small to large polaron
crossover is reported by many experimental groups \cite{bill,lanz}.
There are models \cite {lee,yu2,yar} which incorporate DE 
interaction in a polaronic model. Min and co-workers \cite{lee}
studied the combined model of spin double exchange
and lattice polaron to investigate the effect
of small to large polaron crossover on the magnetic and transport properties
under the mean-field approximation scheme. However, the superexchange 
interaction also plays an important role in manganites and a study 
considering $e$-ph and superexchange interactions in a double exchange 
model is important.    

 In manganites, a competing tendency towards the localization
comes through the $e$-ph interaction which stabilizes a local 
distortion of the
Oxygen octahedron surrounding each Manganese ion.
DE favours the ferromagnetism whereas superexchange
interaction (SE) antiferromagnetism; hence the DE and SE interactions
compete with each other. 
The interplay between lattice-carrier coupling, DE and the SE 
interactions is the reason for the rich phase diagram of the manganites. 
So it would be of great interest to study the role of
SE interaction($J$) in the combined model of DE and $e$-ph interaction. 
In this work we will follow a method based on variational phonon
basis \cite{DS} which is promising for Holstein \cite{DPC,DJC,DJSSC}
and related models \cite{DJC1} to predict the
correct results up to intermediate range 
of hopping ($t$) ($t\le \omega_0$ where $\omega_0$ is the phonon 
frequency) \cite{DJC}. The importance of a variational phonon
basis in predicting accurate results  
has been proved for a two-site system over the entire range of $e$-ph 
couplings \cite{DJC}.
Our previous study \cite{DJC1} on the `two-site' double exchange 
model with a single
polaron as a function of $e$-ph couplings reveals that the 
ferromagnetic (FM) to antiferromagnetic (AFM)
crossover does not coincide always with the 
polaron crossover. If it occurs for some suitable value of 
$J/t$, large changes in the effective hopping and lattice distortion occur  
in the crossover region on application of the magnetic field. 
In ref. \cite{DJC1} a possibility 
of an FM insulating state in between the FM large polaron and AFM 
small polaron state has also been observed  for small values of $J$. 
In this paper 
we consider a many site and many electron double exchange system and
study the role of the superexchange and $e$-ph interaction on the 
magnetic crossover.

The Hubbard Holstein Hamiltonian in presence of double exchange and  
superexchange interaction is given by
\begin{eqnarray}
H &=&  - \sum_{<i,j>,\sigma}
t \cos({\frac{\theta}{2}}) c_{i \sigma}^{\dag} c_{j \sigma} 
+U\sum_{i}n_{i\uparrow}n_{i\downarrow}  
+  \omega_0 \sum_{i}  b_i^{\dag} b_i  
\nonumber \\
&+& J\sum_{<i,j>} \vec S_{i}.\vec S_{j} 
+ g \omega_0  \sum_{i,\sigma}  n_{i \sigma} (b_i + b_i^{\dag}) 
\end{eqnarray} 
where $c_{i\sigma}$ ($c_{i\sigma}^{\dag}$)  
is the annihilation (creation) operator for the electron with spin 
$\sigma$ at site $i$ and $n_{i \sigma}$ is the corresponding number 
operator, $U$ is on-site Coulomb repulsion. 
$b_i$ and $b_{i}^{\dag}$ are the annihilation and 
creation operators, respectively, for the phonons corresponding to 
intramolecular vibrations at site $i$,
$g$ denotes the on-site $e$-ph coupling strength. 
$\vec S_{i}$ and $\vec S_{j}$ are the core-spins at the site $i$ and $j$
respectively, $\theta$ is the angle between the core-spins $\vec S_{i}$ and 
$\vec S_{j}$ and $J$ is the AFM-SE interaction 
between the neighbouring core-spins.
The transfer hopping integral ($t$) is modified to 
$t\cos({\frac{\theta}{2}})$ by the strong Hund's coupling 
between the spins of the core electron and itinerant
electron \cite{zen}. The localized spins are treated classically here.  
It may be mentioned that many workers \cite{rod,lee} have followed
the single orbital approximation on the ground state as
the Jahn-Teller (JT) effect (static and dynamic) will split the $e_g$ double
degeneracy \cite{R} and the mobile $e_g$ electron would occupy the
lower energy orbital at low temperature. The present assumption of
single orbital is expected to be reasonable for doping regime
($x<$0.4-0.5) where CMR occurs. 

For a general value of $e$-ph coupling, the spread and depth of lattice
deformations can be studied using Modified Lang-Firsov(MLF) transformation.
To treat the lattice deformations that produced at different sites around 
the electron variationally we use the MLF transformation \cite{DS},
\begin{equation}
{\tilde H} = e^R H e^{-R}
\end{equation}
\begin{equation}
R=\sum_i \left[\lambda_0 n_i (b_i^{\dag} -b_i)
+\sum_{\delta}\lambda_1 n_i (b_{i+\delta}^{\dag} -b_{i+\delta}) \right]
\end{equation}
with $\lambda_0$ and $\lambda_1$ represent the lattice deformations
at the electron site and next-nearest neighbour sites, respectively. 
When $\lambda_0$=$g$ and $\lambda_1$=$0$,
MLF transformation reduces to Lang-Firsov(LF) transformation
\cite{LF}. At zero temperature the simplest procedure is to make
zero phonon averaging to obtain an effective polaronic Hamiltonian. 
But for a many electron system it is better to use a two-phonon coherent 
state \cite{zheng} for phonon averaging

\begin{equation}
|\psi\rangle_{\rm ph}= \exp{ \left[ \alpha \sum_{i}  (b_i b_i -b_i^{\dag}
 b_i^{\dag}) \right] }|0\rangle_{\rm ph}
\end{equation}
$\alpha$ is the squeezing parameter and treated variationally. 
$\alpha$ is nonzero and its effect becomes significant only for 
intermediate $e$-ph coupling and finite carrier concentration 
\cite{DS}. Increasing 
$\alpha$ enhances the overlapping of the phonon wavefuntions at
the nearest-neighbor sites, hence, increases the polaronic hopping. For
a many electron system, consideration of $\alpha$ lowers the ground 
state energy (for intermediate coupling) and smoothens the polaron 
crossover compared to those obtained by MLF and zero phonon averaging.

 The MLF-transformed Hamiltonian, averaged over the squeezed phonon 
state, yields the effective polaronic Hamiltonian as 
\begin{eqnarray}
\tilde{H}_{\rm eff} &=& \sum_{i,\sigma} \epsilon_p n_{i,\sigma}
 - t_{{\rm p}} \cos({\frac{\theta}{2}})\sum_{<i,j>,\sigma} c_{i \sigma}^{\dag} c_{j \sigma} 
\nonumber\\    
&+& U_{{\rm eff}}\sum_{i}n_{i\uparrow}n_{i\downarrow}  
+ JS^2 \cos{\theta} +V_{1}\sum_{i,j}n_{i}n_{j} 
\nonumber\\    
&+& V_2\sum_{i,\delta+\delta^{'}\ne 0} n_{i+\delta} n_{i+\delta_{'}}
 + N\omega_0 \sinh^2{(2\alpha)}
\end{eqnarray}
where $\epsilon_p$ is the polaronic self-energy, $t_{{\rm p}}$ is
the polaronic hopping, $U_{{\rm eff}}$ is the effective on-site 
interaction, $V_1$ and $V_2$ are interactions between polarons at 
nearest-neighbour and next nearest-neighbour sites, respectively. 
These quantities are obtained as
\begin{eqnarray}
\epsilon_p &=&  - \omega_0~(2 g - \lambda_0)~\lambda_0 
+ z \omega_0\lambda_1^2  \nonumber\\
t_{{\rm p}}&=& t~ \exp{ \left[ \exp{(-4 \alpha)}     
\{-(\lambda_0 - \lambda_1)^2 -(z-1) {\lambda_{1}}^2 \}
  \right] }\nonumber\\    
U_{{\rm eff}} &=& U- 2[  \omega_0~(2 g - \lambda_0)~\lambda_0 
- z \omega_0 {\lambda_{1}}^2 ] \nonumber \\
V_1 &=& - 2( g-\lambda_0) \omega_0  \lambda_1 \nonumber \\
V_2 &=&  \omega_0  \lambda_{1}^2 
\end{eqnarray}
$N$ is the total number of sites in the system.

The ground state energy ($E_G$) of the system per site
is obtained in the framework of Hartree-Fock approximation
\begin{eqnarray}
\frac{E_{G}}{N} &=&  \epsilon_p x_{e} -z ~p~t_{{\rm p}}  \cos({\frac{\theta}{2}}) 
+\frac{U_{{\rm eff}}}{4} x_{e}^2  
+  \frac{J}{2}z S^2   \cos{\theta}
 \nonumber \\
&+&  V_{1} z x_{e}^2 
+ V_2 z^{'} x_{e}^2 
+ \omega_0\sinh^2{(2\alpha)}
\end{eqnarray}
where $z$ is the number of nearest neighbours, $x_e$ is
the number of electrons per site, 
\begin{eqnarray}
x_e &=&\frac{1}{N}\sum_{i} \langle n_i \rangle, ~~~ z^{'}=
\sum_{\delta+\delta^{'}\ne 0}~ 1 \nonumber \\
{\rm and}~ p&=& \langle c_{i,\sigma}^{\dag} c_{j,\sigma} \rangle=
\frac{S_d}{zN}\sum_{\vec{q}} \gamma_{\vec{q}} n_{\vec{q}} \nonumber
\end{eqnarray}
$\gamma_{\vec{q}}=\sum_{j}^{'} e^{i {\vec{q}}. (\vec{R_{i}}-\vec{R_{j}})}$
and  $S_d$ (=2) is the spin degeneracy. For simplicity we choose
a square density of states, then one obtains at zero temperature  
$p=\frac{x_e}{4} (2 -x_e)$ \cite{DS}.
The polaronic variational parameters $\lambda_0$, $\lambda_1$ and 
$\alpha$ as well as the magnetic variational parameter $\theta$ 
are determined from the minimization of the ground state energy. 
For $\theta$ a simple analytical expression is obtained as,
\begin{eqnarray}
\cos{\frac{\theta}{2}}& = & \frac{p ~t_{{\rm p}}}{2JS^2}
 ~~\rm{(for~ nonzero ~solution ~of~} \theta) 
\end{eqnarray}
For $pt_{{\rm p}}$$\geq$$2JS^2$, $\theta$=0 which corresponds to 
FM state. 
In this work we will limit to small values of $t$ ($t < \omega_0$)
so that the error encountered in phonon averaging is small and
does not change our results significantly. We have considered 
lattice deformations only up to the nearest-neighbour 
sites (around the electron) to avoid too many variational 
parameters and because of the fact that 
for $t < \omega_0$ the inclusion of variational parameters 
describing lattice distortions at distant sites 
do not change the qualitative behaviour of the results.
For numerical calculation in this paper we consider 
$z$ and $z^{'}$ are 4 and 12, respectively and energy parameters
$t$, $U$ and $J$ are expressed in unit of $\omega_0$=1.0.

  In Fig. 1 we plot the angle ($\theta$) between the core spins on 
nearest-neighbor sites, the effective hopping 
($t_{{\rm DE}}$ =$t_{{\rm p}}\cos(\theta/2)$) of the
itinerant electron and the on-site lattice deformation ($\lambda_0/g$) 
as a function of $e$-ph coupling ($g$) for different values of  
$J$ and $t$. Figs. 1(a) and 1(b) (for $t$=0.7) show that for small $g$ 
the ground state is the FM with large polaron as carriers. The 
signature of large polaron is evident from weak polaronic 
reduction in hopping and the reduced value of $\lambda_0/g$. 
For large $g$ the ground state is the AFM state ($\theta$$=$$\pi$) 
with small polarons as carriers. The above result may be understood 
considering the fact that 
increasing $e$-ph coupling leads to a polaronic reduction of the 
kinetic energy which in turn destroys the FM state since the 
stability of the FM state requires $pt_p \geq 2JS^2$. Similar behavior 
has also been obtained by recent numerical studies \cite{Feh}.
Fig. 1 shows that the magnetic transition occurs simultaneously
with the large to small polaron crossover within the present model. 
With increasing $J$ the magnetic transition as well as
the polaron crossover shifts towards lower values of $e$-ph
coupling and becomes smoother. Similar effect is also obtained
by decreasing the value of $t$ (Figs. 1(c) and 1(d)).   
For $t$=0.5 and higher values of $J$ ($JS^2$=0.1) the canted magnetic 
state is stable for weak $e$-ph coupling and the transition from the 
canted state to the AFM state with increasing $g$ is very smooth. 
Fig. 1 also shows that the nature of the magnetic transition 
depends on both $t$ and $J$. 
\begin{center}
\begin{figure}
\centering
\includegraphics[scale = 0.38,angle=0]{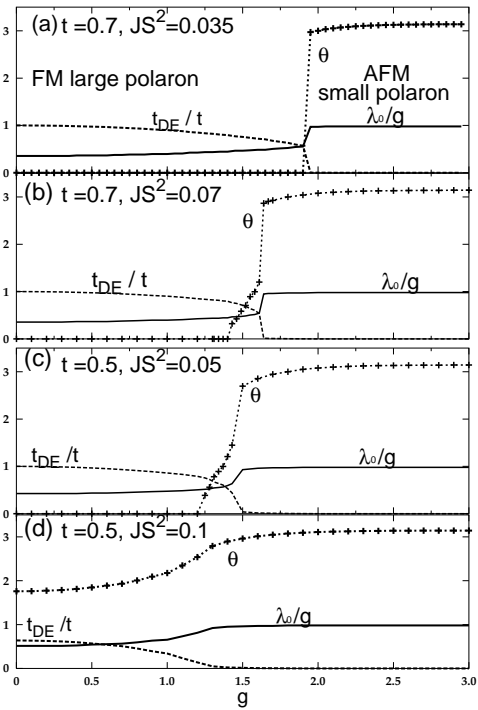}
\caption{Plot of 
$\lambda_0/g$, $t_{{\rm DE}}/t$ and $\theta$ (in radian)
for $x_e$=0.3, $U$=1.0 as a function of $g$. (a) $t$=0.7 and
$JS^2$=0.035; (b) $t$=0.7 and  $JS^2$=0.07; 
(c) $t$=0.5 and $JS^2$=0.05 and (d) $t$=0.5 and $JS^2$=0.1. 
}
\end{figure}
\end{center}
\begin{center}
\begin{figure}[t]
\centering
\includegraphics[scale = 0.34,angle=270]{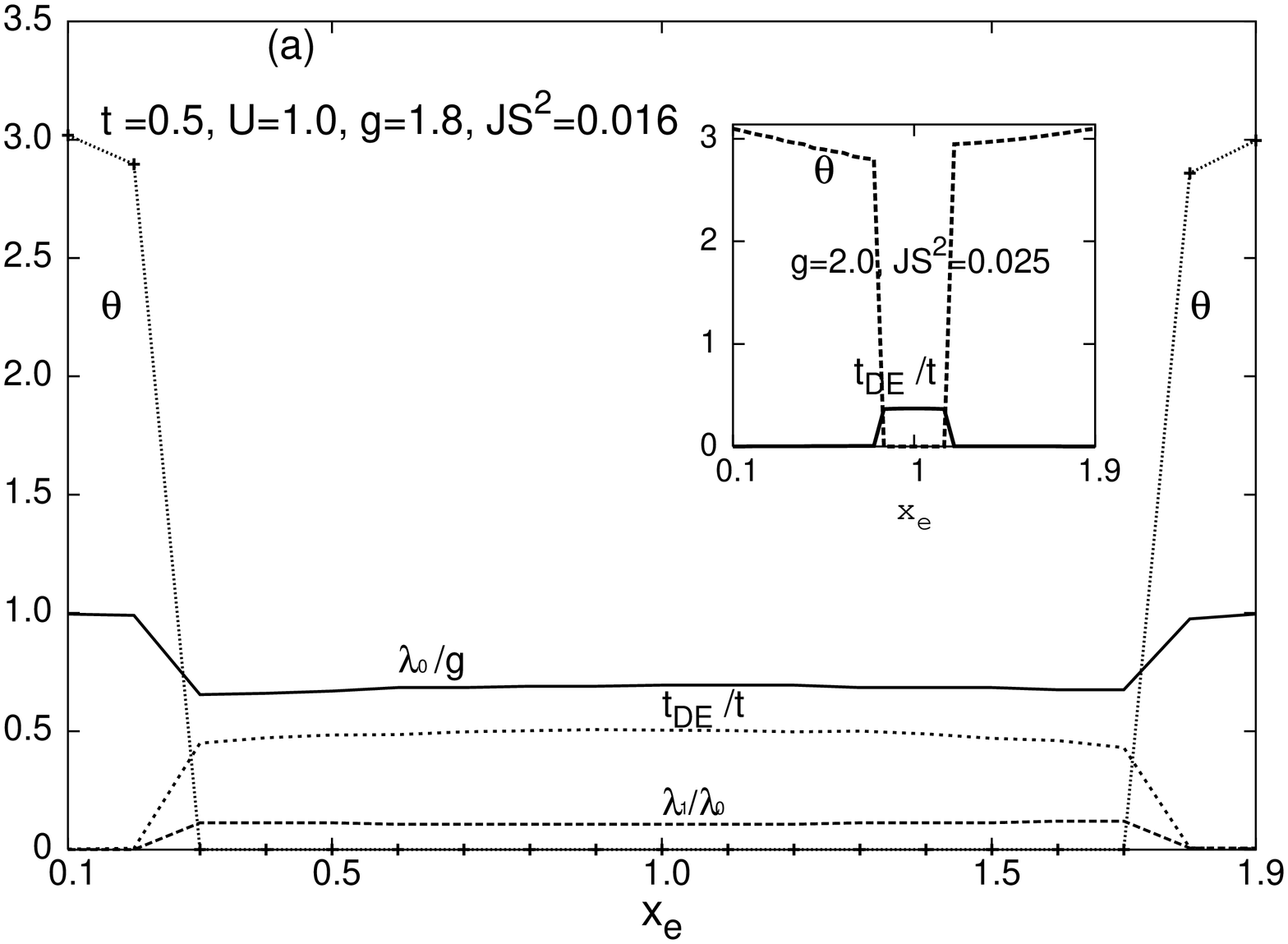}
\includegraphics[scale = 0.34,angle=270]{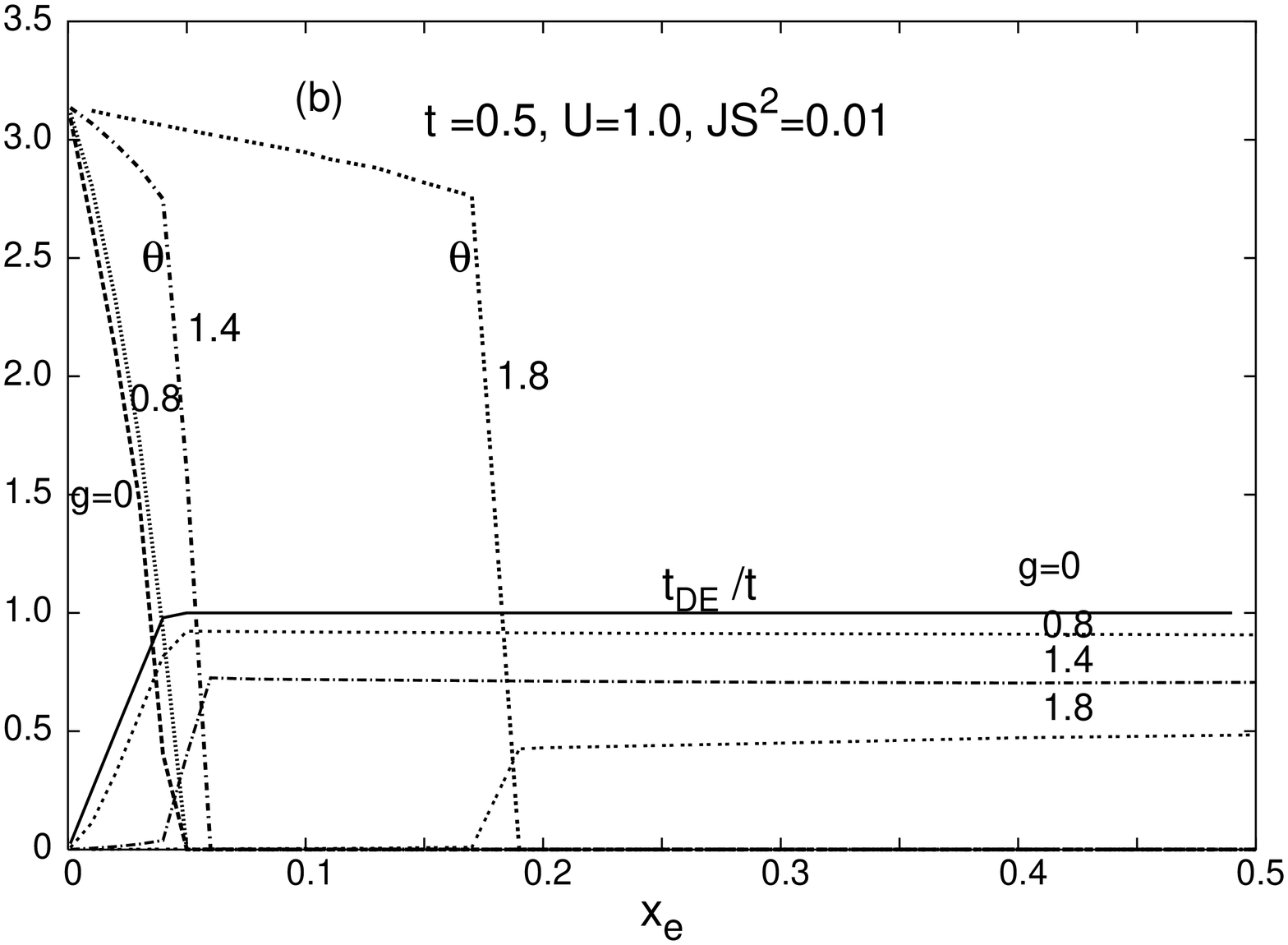}
\caption{
(a) Plot of $\theta$, $t_{{\rm DE}}/t$, $\lambda_0/g$
and $\lambda_1/\lambda_0$  as a function of electron 
concentration ($x_e$) for $t$=0.5, $U$=1.0, $ JS^2$=0.016  and 
$g$=1.8. Inset figure for higher values 
of $ JS^2$ and $g$. 
(b) Plot of $\theta$, $t_{{\rm DE}}/t$ vs. $x_e$ with $t$=0.5, 
$ JS^2$=0.01 for different values of $e$-ph coupling.
}
\end{figure}
\end{center}

 In Fig. 2(a) the variations of $\lambda_0$, $\lambda_1$, $t_{{\rm DE}} $ and
$\theta$ with electron concentration ($x_e$) are shown
for different values of $g$ and $J$.
For low electron concentrations, the ground state is AFM and the carriers
are small polarons. As the electron concentration increases 
$\lambda_0$ decreases while $\lambda_1$ increases depicting a small to
large polaron crossover. For intermediate electron concentrations 
the ground state is FM with large polaronic carriers. The 
Hamiltonian (1), that we considered, has a particle-hole symmetry. 
This is reflected in Fig. 2(a) where all the physical quantities (presented
in the figure) are symmetric with respect to half filling ($x_e$=1). 
In Fig. 2(b) we have plotted $\theta$ and $t_{DE}$ against $x_e$ 
for different values of $e$-ph coupling. In absence of $e$-ph 
coupling ($g$=0) spin canting ($0<\theta<\pi$) 
starts at the smallest carrier concentration
and the canting increases with $x_e$ until the material becomes
ferromagnetic at a low value of critical concentration $x_{cr}$ 
\cite{Khomskii}.   
With increasing $g$ the  AFM small polaronic state remains the ground
state for a range of low carrier concentration and the value of
$x_{cr}$ increases. For $JS^2/t$=0.02, which is reasonable 
for manganites, the AFM-FM transition occurs around $x_e$$\sim$0.2 
for $g$=1.8. Fig. 2(b) shows that 
the region of AFM small polaron state extends at the expense of
the FM large polaronic region with increasing
$J$ or $g$ (also evident in the inset of Fig. 2(a)). This is
simply due to the fact that the FM large polaron state is destroyed 
by increasing the AFM interaction as well as by the suppression
of the kinetic energy with increase of  $e$-ph coupling.  
Hence the combined role of $e$-ph coupling and SE interaction
determines the physics of the system. 

 It may be mentioned that near half filling ($x_e$$\sim$1) the
system is susceptible to show charge ordering depending on the 
choice of the parameters. However, we have not addressed this issue here,
because our main interest in this paper is
in the region from low to intermediate filling where
a transition from an AFM small polaronic state to FM large polaron 
state occurs as the density of the carriers increases and the 
$e$-ph coupling plays a crucial role in this transition.

\begin{center}
\begin{figure}
\centering
\includegraphics[scale = 0.34,angle=270]{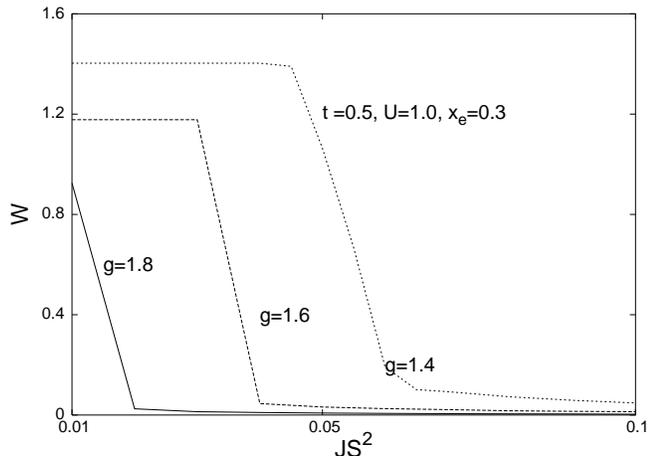}
\caption{
Bandwidth ($W= z t_{\rm p} \cos{\theta/2}$)
as a function of $JS^2$ for different values of $g$
with $t$=0.5, $U$=1.0 and $x_e$=0.3. 
}
\end{figure}
\end{center}

 In manganites there are SE interactions between nearest neighbor
$e_g$ electrons. This interaction originates from the virtual
hopping of $e_g$ electrons and is proportional to the square of the 
hopping matrix element and inversely proportional to 
the increase in energy ($\sim$ on-site Coulomb repulsion $U_{\rm eff}$)
in the intermediate state when two electrons occupy the same site. 
The SE interaction between the $e_g$ electrons determines the 
orbital ordering temperature in undoped manganite \cite{Maekawa}.
A relevant question may be asked whether this SE interaction will
be reduced by the small polaron formation as the polaronic hopping is 
suppressed. One of us \cite{Das} has shown earlier in the context 
of Hubbard 
model that the SE interaction is not reduced by the polaron formation
because in virtual hopping the lattice deformations are not transferred
from site to site. Rather a slight decrease in the value of
the on-site repulsion  due to polaron formation 
may slightly increase the SE interaction between $e_g$ electrons.

  In Fig.3 we plot the electronic effective half-bandwidth ($W$) 
as a function of the AFM interaction for different (intermediate) 
values of e-ph coupling. It is seen that the bandwidth has a unique
dependence on $J$ also. In the metallic FM region the bandwidth is 
independent of $J$; whereas in the canted state 
the bandwidth is suppressed drastically with increasing $J$.
Narrowing of the electronic bandwidth
is considered to be one of the origins of the suppression effect of
the transition temperature ($T_c$) for doped manganites. Strong $e$-ph
interaction favours the band narrowing due to the small polaron formation.
The AFM-SE interaction may be considered as another candidate
for band narrowing mechanism. Our mean-field result is qualitatively
consistent with the Monte Carlo study of FM Kondo lattice model \cite{Yi}.
Fig. 3 also shows that increasing $e$-ph coupling extends the 
AFM small polaron phase with negligible bandwidth in the 
$JS^2$ space, similar to that observed as a function of electron
concentration in Fig. 2(a).  

  In summary, we have studied the role of SE interaction ($J$)
in a many-site DE-Holstein model within the mean-field
approximations using the variational phonon basis.  
The polaron crossover and the magnetic transition
are studied as a function of $e$-ph coupling ($g$) and electron
concentration for different values of $J$. The suppression
of the FM phase and the narrowing of the bandwidth with both $J$ 
and $g$ are observed. The small to large polaron
crossover as well as the AFM-FM transition is found to occur 
at a higher electronic concentration as $J$ or $g$ increases.

Electronic address for correspondence: jayita@physics.postech.ac.kr

\end{document}